\documentclass[prb,twocolumn,superscriptaddress,showpacs,preprintnumbers,amsmath,amssymb,floats]
{revtex4}

\usepackage{txfonts}
\usepackage{amssymb}
\usepackage{graphicx}
\usepackage[dvips]{color}

\begin{document}
\title{Tuning the electronic structure of Sr$_2$IrO$_4$ thin films by bulk electronic doping using molecular beam epitaxy}

\author{M. Y. Li}
\author{Z. T. Liu}
\author{H. F. Yang}
\author{J. L. Zhao}
\affiliation{State Key Laboratory of Functional Materials for Informatics, Shanghai Institute of Microsystem and Information Technology (SIMIT), Chinese Academy of Sciences, Shanghai 200050, China}

\author{Q. Yao}
\affiliation{State Key Laboratory of Functional Materials for Informatics, Shanghai Institute of Microsystem and Information Technology (SIMIT), Chinese Academy of Sciences, Shanghai 200050, China}
\affiliation{State Key Laboratory of Surface Physics, Department of Physics,
and Advanced Materials Laboratory, Fudan University, Shanghai 200433, China}

\author{C. C. Fan}
\author{J. S. Liu}
\author{B. Gao}
\author{D. W. Shen}\email{dwshen@mail.sim.ac.cn}
\author{X. M. Xie}
\affiliation{State Key Laboratory of Functional Materials for Informatics, Shanghai Institute of Microsystem and Information Technology (SIMIT), Chinese Academy of Sciences, Shanghai 200050, China} 

\begin{abstract}
By means of oxide molecular beam epitaxy with shutter-growth mode, we have fabricated a series of electron-doped (Sr$_{1-x}$La$_x$)$_2$IrO$_4$ (001) ($x$ = 0, 0.05, 0.1 and 0.15) single crystalline thin films and then investigated the doping dependence of electronic structure utilizing in-situ angle-resolved photoemission spectroscopy. We find that with increasing doping proportion, the Fermi levels of samples progressively shift upward. Prominently, an extra electron pocket crossing the Fermi level around the M point has been evidently observed in 15\% nominal doping sample. Moreover, bulk-sensitive transport measurements confirm that doping effectively suppresses the insulating state with respect to the as-grown Sr$_2$IrO$_4$, though doped samples still remain insulating at low temperatures due to the localization effect possibly stemming from disorders including oxygen deficiencies. Our work provides another feasible doping method to tune electronic structure of Sr$_2$IrO$_4$.

\end{abstract}


\maketitle

5$d$ transition-metal oxides (TMOs) have recently drawn a lot of attention. In these compounds, the complicated interplay between spin-orbit coupling (SOC) and electron correlations has been suggested to host multiple novel quantum states, including topological Mott insulators~\cite{TMI}, Weyl semimetals~\cite{WS_1,WS_2}, axion insulators~\cite{axion_1,axion_2}, and spin liquids~\cite{spinL_1,spinL_2,spinL_3}. These studies were mostly initialized by the pioneering experiments on the prototype layered perovskite Sr$_2$IrO$_4$, in which the strong SOC was revealed to lift the orbital degeneracy and then result in a narrow half-filled $J_{eff}$ = 1/2 band that even the relatively weak Coulomb repulsion of 5$d$ electrons could induce a Mott metal-insulator transition (MIT) therein~\cite{5dMIT_1,5dMIT_2}. Scattering experiments discovered that, this Mott insulator, showing an effective pseudospin 1/2 antiferromagnetic (AFM) order at low temperature, can be well described by the Heisenberg model with an exchange coupling of 60 to 100 meV~\cite{psHeisbg_1,psHeisbg_2}. Such findings indicate that the low-energy behavior of this single-layer iridate rather resembles that of cuprates. Consequently, it is tempting to investigate the carrier doping of Sr$_2$IrO$_4$, which might pave the way to discover a new family of unconventional superconductors. Indeed, a recent theoretical work has predicted the unconventional superconductivity in the electron-doped Sr$_2$IrO$_4$~\cite{theorySC}.

In this context, various experimental attempts have been performed to achieve the effective carrier doping of Sr$_2$IrO$_4$~\cite{doping_1,doping_2,doping_3,doping_4,doping_5,doping_6,doping_7}. So far, one of the most promising progresses is the try of surface electron doping of Sr$_2$IrO$_4$ via \emph{in-situ} potassium deposition, in which Fermi arcs and pseudogap behavior have been reported~\cite{Fermiarc}; nevertheless, this finding is not consistent with the reports on the sibling La-doped Sr$_3$Ir$_2$O$_7$, in which only small Fermi pockets rather than Fermi arcs were revealed~\cite{327dope_1,327dope_2}. This controversy naturally raises one open question, namely, whether the existence of Fermi arcs is an universal property of the electron-doped iridates irrespective of the specific doping method. As there has been no equivalent study reported so far for La-doped Sr$_2$IrO$_4$, this important question still remains unanswered.

In this letter, we synthesize a series of high-quality La-doped Sr$_2$IrO$_4$ thin films on SrTiO$_3$(001) single crystal substrates by means of oxide molecular beam epitaxy (OMBE), by which the metastable phases with large electron doping levels could be stabilized in the formation of thin films. We can thus investigate the electron-doping dependence of electronic structure of bulk Sr$_2$IrO$_4$ through \emph{in-situ} angle-resolved photoemission spectroscopy (ARPES). We find that, with increasing the La-doped proportion, the Fermi levels of samples gradually shift upward to the conduction band, which is consistent with the electron doping. In addition, we recognize an extra electron-like band crossing the Fermi level upon 15\% nominal La-doping, forming one pocket-like feature around the boundary of the reduced Brillouin zone (BZ), in stark contrast to as-grown Sr$_2$IrO$_4$ samples. Although electric transport measurements manifest the lanthanum substitution of Sr can significantly suppress the resistivity of this system, the doped Sr$_2$IrO$_4$ still show clear insulating behavior. We argue that this may be caused by the random carrier hoping between localized states in our system due to inevitable oxygen related defects.

\begin{figure}[t]
\includegraphics[width=8.5cm]{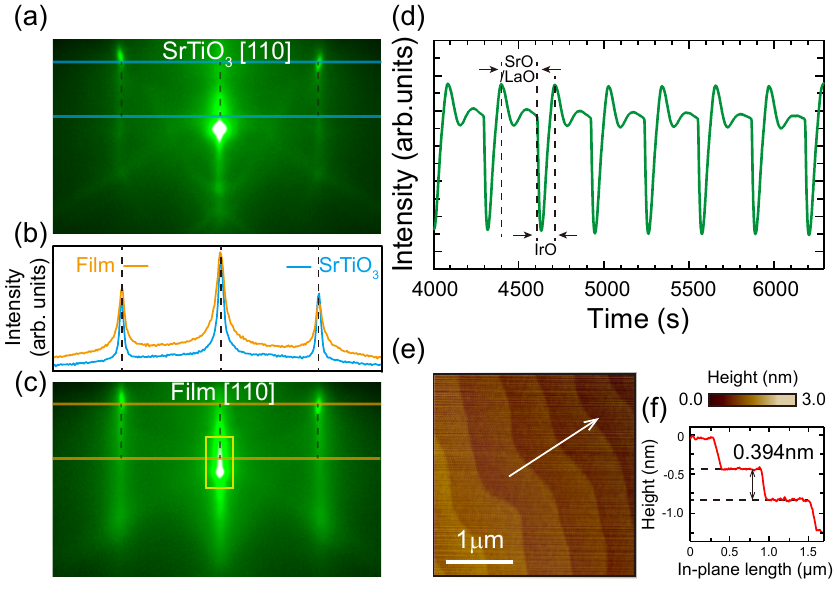}\\
\caption{(Color online) (a) (c) RHEED patterns of SrTiO$_3$(001) substrate and as-grown (Sr$_{0.95}$La$_{0.05}$)$_2$IrO$_4$ along [110]$_p$ azimuth direction, respectively. (b) RHEED intensity curves integrated within the rectangle windows in panels (a) and (c). (d) Time dependence of the intensity of [00] diffraction streak in RHEED pattern of (Sr$_{1-x}$La$_x$)$_2$IrO$_4$(x = 0.05) thin film during deposition. (e) AFM topographic image of as-grown (Sr$_{0.95}$La$_{0.05}$)$_2$IrO$_4$ thin film. (f) The in-plane length dependence of height along the arrow direction in panel (e).}
\label{tabular}
\end{figure}

The La-doped Sr$_2$IrO$_4$(001) films were deposited onto SrTiO$_3$(001) single crystalline substrates using a DCA R450 OMBE system. The flat stepped surface of substrates was obtained by etching in hydrofluoric acid buffer solution (BHF) and annealing in a muffle furnace according the standard precedure~\cite{BHF}. The in-plane lattice constant of SrTiO$_3$ substrates is 3.905 {\AA}, which leads to a 0.44\% tensile strain to the pseudo-cubic (La,Sr)$_2$IrO$_4$ (3.888 {\AA}). The shuttered growth mode was applied to synthesize films in the distilled ozone atmosphere of 2 $\times$ 10$^{-6}$ Torr. During the growth, the temperature of substrates was kept at 800 $^\circ$C verified by the thermocouple behind the sample stage. Moreover, the overall growth rate and surface structure of thin films were monitored by \emph{in-situ} reflection high-energy electron diffraction (RHEED) during growth. Strontium (lanthanum) and iridium were evaporated from Knudsen effusion cells and an electron beam evaporator, respectively. All the doping levels as mentioned in this article are nominal values.

Figures~\ref{tabular} (a) and (c) show the RHEED patterns of SrTiO$_3$(001) substrate before the growth and after the growth of 6 unit-cell (Sr$_{0.95}$La$_{0.05}$)$_2$IrO$_4$ thin film, respectively, which were both taken with a glancing electron beam parallel to the [110]$_p$ azimuthal direction. These RHEED images exhibit prominent Kikuchi lines, indicating the high crystalline perfection of films. We compared the RHEED intensity curves integrated within the rectangle windows of the (Sr$_{0.95}$La$_{0.05}$)$_2$IrO$_4$ film and substrate [Fig.~\ref{tabular}(b)], and the negligible shift of the intensity peak indicates there is no evident lattice relaxation for the strained film. The amplitudes and periods of RHEED intensity oscillations of the [00] diffraction streak tend to be constant once the Sr(La)/Ir flux ratio is adjusted to be close to stoichiometric 2. However, to further confirm that the shuttered SrO/(La$_2$O$_3$)$_{0.5}$ and IrO$_2$ monolayers are complete during the growth, we measured the total thickness of films and their $c$-axis lattice constants through X-ray diffraction (XRD) to estimate the actual number of unit cells. By comparing this value to the number of shutting periods of our calibration sample, we can then deduce the scaling factor, by which we can calibrate the Sr/La and Ir shutter opening times. In this way, the layer-by-layer growth of (Sr$_{1-x}$La$_{x}$)$_2$IrO$_4$ films could be achieved.

\begin{figure}[t]
\includegraphics[width=8.5cm]{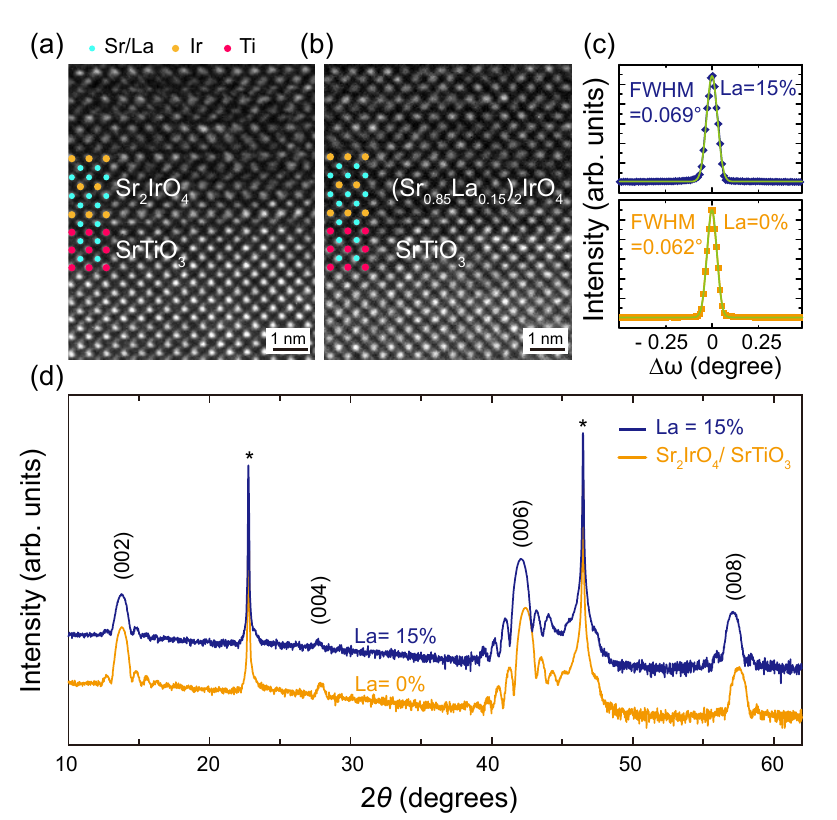}\\
\caption{(Color online) (a), (b) TEM cross-sections of Sr$_2$IrO$_4$ and (Sr$_{0.85}$La$_{0.15}$)$_2$IrO$_4$ thin films on SrTiO$_3$(001) substrates, respectively. (c) Rocking curves of (006) diffraction peaks of Sr$_2$IrO$_4$ (orange symbol-line) and (Sr$_{0.85}$La$_{0.15}$)$_2$IrO$_4$ (blue symbol-line) thin films, respectively. Green lines are Gauss curves for fitting the rocking curves of thin films. (d) X-ray diffraction $\theta$-2$\theta$ scans of 8-nm-thick (Sr$_{1-x}$La$_x$)$_2$IrO$_4$ (x = 0 and 0.15) thins films on SrTiO$_3$(001) substrates. The characteristic peaks of substrates are labeled with an asterisk.}
\label{structure}
\end{figure}

\begin{figure*}[t]
\includegraphics[width=17.5cm]{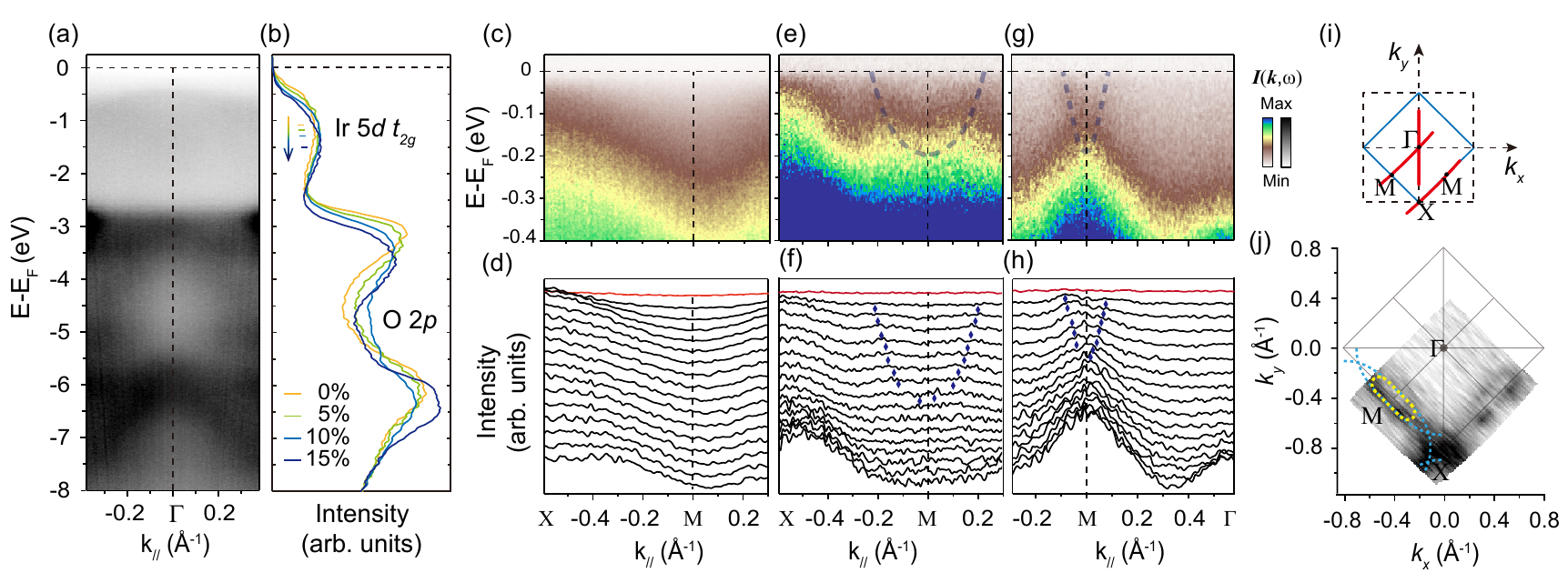}\\
\caption{(Color online) (a) Photoemission intensity plots of Sr$_2$IrO$_4$ thin film along $X$-$\Gamma$-$X$ direction, taken with 21.2 eV photon energy at 40 K. (b) The valence-band photoemission spectra at $\Gamma$ point for (Sr$_{1-x}$La$_x$)$_2$IrO$_4$ thins films. (c) Photoemission intensity plots along $X$-$M$ of Sr$_2$IrO$_4$ thin film. (e), (g) Photoemission intensity plots along $X$-$M$ and $M$-$\Gamma$ of (Sr$_{0.85}$La$_{0.15}$)$_2$IrO$_4$, respectively. (d), (f), (h) The momentum distribution curves for the data in panels (c) , (e) and (g). (i) The two-dimensional projection of the Brillouin zone and the high-symmetry directions. The blue solid lines represent the reduced Brillouin zone. (j) Fermi surface map integrated over [$E_F$ - 15~meV, $E_F$ + 15~meV] of (Sr$_{0.85}$La$_{0.15}$)$_2$IrO$_4$ thin film. Yellow dashed lines guide for the elliptical Fermi surface around M. BLue dashed lines illustrate the reflected shadow band of Fermi arcs. All the APRES data were taken with 21.2 eV photon energy at 40 K.}\label{ARPES}
\end{figure*}

Figure~\ref{tabular}(d) shows a typical RHEED intensity oscillation of the [00] diffraction streak (surrounded by a yellow pane in Fig.~\ref{tabular}(c)) of such a calibrated film. By means of the \emph{ex-situ} atomic force microscopy (AFM), we further examined the typical surface morphology of our thin film, as displayed in Fig.~\ref{tabular}(e). The film surface shows a well defined terrace-step structure with a typical step height of 0.394 nm [Fig.~\ref{tabular}(f)], preserving the flat terraces and surface morphology of SrTiO$_3$(001) substrates. This finding confirms the layer-by-layer growth mode of our films. Moreover, since all growths of our films were terminated after an integral number of oscillation finishing, we can expect a single IrO$_2$ surface termination for all the films.

To characterize the phase purity and crystallinity of (Sr$_{1-x}$La$_x$)$_2$IrO$_4$ thin films, the high-resolution cross-sectional transmission electron microscopys (TEM) and the \emph{ex-situ} XRD measurements were performed. Figs. 2(a) and (b) show the TEM images of the typical Sr$_2$IrO$_4$ and (Sr$_{0.85}$La$_{0.15}$)$_2$IrO$_4$ thin films deposited on SrTiO$_3$(001) substrates. Both data demonstrate the continuous and evenly spaced SrO/(La$_2$O$_3$)$_{0.5}$ and IrO$_2$ layers and the atomically sharp interfaces between the epitaxial thin films and the substrates, with no discernible interdiffusion or vacancy observable. The XRD results of typical (Sr$_{1-x}$La$_x$)$_2$IrO$_4$ thin films are shown in Figs. ~\ref{structure}(c) and (d). The $\theta$-2$\theta$ scans of the films [Fig.~\ref{structure}(d)] are consistent with the growth of phase-pure (001)-oriented (Sr$_{1-x}$La$_x$)$_2$IrO$_4$. The clear Kiessig interface fringes indicate smooth film surfaces and sharp interfaces between the thin films and substrates. We also measured the rocking curves around the (006) peaks of Sr$_2$IrO$_4$ and (Sr$_{0.85}$La$_{0.15}$)$_2$IrO$_4$, as shown in Fig.~\ref{structure}(c). The full widths at half maximum (FWHM) of these curves are 0.062$^{\circ}$ and 0.069$^{\circ}$, respectively, indicating the good epitaxial film quality. In addition, through Bragg's function on the (006) peak positions, we can deduce the out-of-plane \emph{c} lattice constants of  Sr$_2$IrO$_4$ and (Sr$_{0.85}$La$_{0.15}$)$_2$IrO$_4$ to be 12.782{\AA} and 12.821{\AA}, respectively. For comparison, the out-of-plane \emph{c} lattice constant of bulk Sr$_2$IrO$_4$ was determined to be 12.899{\AA}~\cite{214sturcture}. Obviously, the in-plane tensile strain for the epitaxial films results in the compression in the out-of-plane $c$ direction, while the doping of lanthanum with smaller ionic radius partially counterbalances this effect. From another perspective, this finding confirms the real bulk doping of lanthanum in our (Sr$_{1-x}$La$_x$)$_2$IrO$_4$ films grown by OMBE.

After growth, we performed \emph{in-situ} ARPES measurements on these samples in order to investigate the electron doping evolution of (Sr$_{1-x}$La$_x$)$_2$IrO$_4$. Thin films were transferred through an ultrahigh vacuum buffer chamber (1.0 $\times$ 10$^{-10}$ torr) to the combined ARPES chamber for measurements immediately after the growth. This ARPES system is equipped with a VG-Scienta R8000 electron analyzer and a SPECS UVLS helium discharging lamp. The data were collected at 40~K under ultrahigh vacuum of 8 $\times$ 10$^{-11}$ torr. The angular resolution was 0.3 $^{\circ}$, and the overall energy resolution was set to 15~meV (HeI, 21.2~eV photon energy). During the measurements, the films were stable and did not show any sign of degradation.

Fig.~\ref{ARPES}(a) shows the valence band photoemission intensity plot for Sr$_2$IrO$_4$ epitaxial films along $X$-$\Gamma$-$X$ high-symmetry direction highlighted by the red line in Fig. 3(i). By comparing with DFT calculations, we can identify that the features between - 7.0 and - 2.0 eV are mainly contributed by the O $2p$ states, while the Ir $t_{2g}$ orbitals mainly distribute the feature located at around - 0.5~eV, as illustrated by the integrated spectra in momenta around $\Gamma$ point [Fig.~\ref{ARPES}(b)]~\cite{5dMIT_2,214band_1,214band_2,214band_3}. While, upon further La doping (from $x$ = 0 to 0.15 nominal doping level), we discovered a continuous energy shift (as large as~500 meV) for these spectral features as shown in Fig.~\ref{ARPES}(b). This finding confirms the gradual electron doping into the insulating Sr$_2$IrO$_4$ by substituting Sr with La, which would increase the chemical potential of this system and keep pushing valence band (VB) features to higher binding energy. Here, all spectra have been renormalized by the spectral intensity at the binding energy of 8.0 eV for comparison.

Surprisingly, as for the sample with nominal 15\% La-doping, we observe an extra electron-like band crossing the Fermi level[Figs.~\ref{ARPES}(e) and (g)], while the non-doped Sr$_2$IrO$_4$ just presents as a typical insulator with an energy gap[Fig.~\ref{ARPES}(c)]. Moreover, the comparison between the corresponding momentum distribution curves in Figs.~\ref{ARPES}(d), (f) and (h) further confirms this finding. As shown in Fig.~\ref{ARPES}(j), the band forms one elliptical electron pocket around the $M$ point of the reduced Brillouin zone which is guided by yellow dashed line[Fig.~\ref{ARPES}(i)]. It seems that our findings are in qualitative agreement with the Fermi pockets discovered in the metallic electron-doped Sr$_3$Ir$_2$O$_7$, in which electron carriers are doped into the conduction band and produces a small Fermi surface containing only the added $x$ electrons~\cite{327dope_2}. However, we could still not simply rule out the possibility of the Fermi arcs. Taking into account of the significant octahedral rotation and the possible antiferromagnetic (AF) ordering in (Sr$_{1-x}$La$_x$)$_2$IrO$_4$, it is highly possible for ARPES to observe the reflected shadow bands of the Fermi arcs, which would as well produce the pocket-like feature as illustrated by the blue dashed lines in Fig.~\ref{ARPES}(j). To further pin down the origin of this pocket-like feature, a more detailed doping dependence of the low-lying electronic structure is highly imperative.

\begin{figure}[t]
\includegraphics[width=8.5cm]{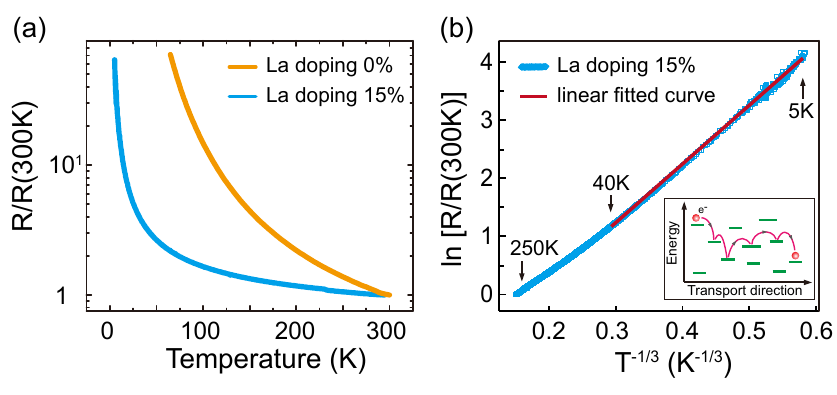}\\
\caption{(Color online)(a) The temperature dependence of resistance for (Sr$_{1-x}$,La$_x$)$_2$IrO$_4$(001) (x = 0 and 0.15) epitaxial films, normalized by the resistance at 300 K temperature. (b) The logarithm of the in-plane normalized resistivity as a function of T$^{-1/3}$ for (Sr$_{0.85}$,La$_{0.15}$)$_2$IrO$_4$ epitaxial film. Inset: Sketch of a typical two-dimensional variable range hoping model.}
\label{transport}
\end{figure}

To further determine whether our findings are caused by the surface or the bulk electronic structure, we compared the in-plane resistivity of films with different doping levels. As shown in Fig.~\ref{transport}(a), for the as-grown Sr$_2$IrO$_4$, the normalized R-T curve manifests a typical insulator behavior and the resistivity has been out of measurement range when the temperature is below 70 K; nevertheless, though the film with 15\% nominal doping level still behaves as an insulator, its in-plane resistivity has dropped by around three orders at 70 K, and it is still in the range of our measuring system at as low as 5 K. This prominent change of the bulk properties of films is in good agreement with our ARPES result, and it undoubtedly supports that our findings are mainly caused by the electron doping into the bulk Sr$_2$IrO$_4$. We note that (Sr$_{0.85}$La$_{0.15}$)$_2$IrO$_4$ is still insulating though there is finite spectra weight in the vicinity of Fermi energy ($E_F$), which implies that this system might be in the strongly localized regime~\cite{RTlocalized}. Fig.~\ref{transport}(b) shows the logarithm of the in-plane resistivity as a function of T$^{-1/3}$ for a temperature range from 5 to 300 K. In the low temperature range (5 to 40 K), we could observe an excellent agreement with a linear fit, consistent with the typical two-dimensional variable range hopping (VRH) model~\cite{VRH}. In this case, carriers would hop between localized states as sketched by the inset of Fig.~\ref{transport}(b). At higher temperatures, though there exists a slight deviation from the linear fit, we can conclude that the VRH may still be the main conduction mechanism. Recently, one scanning tunneling microscopy and spectroscopy experiment discovered some oxygen related defects on the surface of Sr$_2$IrO$_4$, which are likely native to the sample and regardless of the good sample quality~\cite{oxygendefect}. Such random distribution of the defects are in good agreement with the VRH behavior in our in-plane resistivity data.

In summary, we have grown a series of La-doped (Sr$_{1-x}$La$_x$)$_2$IrO$_4$(001) ($x$ = 0, 0.05, 0.1, and 0.15) epitaxial films on SrTiO$_3$(001) substrates and studied the corresponding electronic structure and transport properties. By increasing the La-doping level, Fermi level for the samples are drifted up getting close to higher binding energy. Moreover, a fast dispersing state is observed around $M$ point for the samples with 15\% La-doping. In addition, the R-T curves of (Sr$_{0.85}$La$_{0.15}$)$_2$IrO$_4$ thin film is suppressed under the one of undoped Sr$_2$IrO$_4$ thin film. Although the La-doped proportion is as high as 15\%, the films still remain insulators, which might be resulted from the strong localized effect induced by disorders including the oxygen defects in the samples.

We gratefully acknowledge the helpful discussion with Prof. Donglai Feng, Dr. Rui Peng and Dr. Wei Li. This work was supported by National Basic Research Program of China (973 Program) under the grant Nos. 2011CBA00106 and 2012CB927400, the National Science Foundation of China under Grant Nos. 11274332 and 11227902, and Helmholtz Association through the Virtual Institute for Topological Insulators (VITI). M. Y. Li and D. W. Shen are also supported by the "Strategic Priority Research Program (B)" of the Chinese Academy of Sciences (Grant No. XDB04040300).

\end{document}